\documentclass[aps,pra,preprint,superscriptaddress,longbibliography]{revtex4-2}

\usepackage{graphicx}
\usepackage{amsmath}
\usepackage{amsfonts}
\usepackage{amssymb}
\usepackage{color}
\usepackage[pdftex]{hyperref} 

\begin{document}

 \title{A supersymmetry journey from the Jaynes--Cummings to the anisotropic Rabi model}

    \author{A. Kafuri} 
    \email[email: ]{anuar.kafuri@cinvestav.mx}
    \affiliation{Departamento de Física, Cinvestav, Avenida Instituto Politécnico Nacional 2508, San Pedro Zacatenco, 07360 Gustavo A. Madero, Ciudad de México, Mexico}

    \author{F.~H. Maldonado-Villamizar} 
    \email[email: ]{fmaldonado@inaoep.mx}
    \affiliation{CONAHCYT - Instituto Nacional de Astrof\'isica, \'Optica y Electr\'onica, Calle Luis Enrique Erro No. 1. Sta. Ma. Tonantzintla, Pue. C.P. 72840, Mexico}

    \author{A. Moroz} 
    \email[email: ]{wavescattering@yahoo.com}
    \affiliation{Wave-scattering.com}

    \author{B.~M. Rodr\'iguez-Lara}
    \email[e-mail: ]{bmlara@upp.edu.mx}
    \affiliation{Universidad Polit\'ecnica de Pachuca. Carr. Pachuca-Cd. Sahag\'un Km.20, Ex-Hda. Santa B\'arbara. Zempoala, 43830 Hidalgo, Mexico }

\begin{abstract}
We revisit the Jaynes--Cummings and anti-Jaynes--Cummings model through the lens of Lie theory, aiming to highlight the efficacy of an operator-based approach for diagonalization.
We focus on explicitly delineating the steps to go from an underlying abstract supersymmetry, provided by the $u(1 \vert 1)$ superalgebra, into concrete proper states and energies in the laboratory frame.
Additionally, we explore the anisotropic Rabi model possessing an underlying supersymmetry, provided by the $osp(2 \vert 2)$ superalgebra, in a squeezed reference frame where it is possible to approximate its spectral characteristics by an effective Jaynes--Cummings model.
Finally, we identify a regime for a factorizable anisotropic Rabi model, exhibiting an equally spaced, double degenerate energy spectrum with a unique ground state energy.
Our work aims to merge mathematical physics with practical quantum optics, underscoring the critical role of Lie theory. 
\end{abstract}

\maketitle
\newpage

\section{Introduction}
The Jaynes--Cummings (JC) model \cite{JaynesCummings1963}, 
\begin{align}
     \frac{\hat{H}}{\hbar} =&~ \omega \hat{a}^{\dagger} \hat{a} + \frac{1}{2} \omega_{0} \hat{\sigma}_{z}  + \lambda \left( \hat{a} \hat{\sigma}_{+} + \hat{a}^{\dagger} \hat{\sigma}_{-} \right), 
\end{align}
is a cornerstone in quantum optics providing a fundamental understanding of the interaction between quantum electromagnetic radiation, described by a boson mode characterized by frequency $\omega$ and annihilation (creation) operators $\hat{a}$ ($\hat{a}^{\dagger}$), with matter, described by a spin-half fermion mode characterized by frequency $\omega_{0}$ and annihilation (creation) operators $\hat{\sigma}_{-}$ ($\hat{\sigma}_{+}$), with a real coupling strength $\lambda$.
Despite its inconspicuous Hamiltonian form, its algebraic richness and complexity emerges in the form of the Lie superalgebra $u(1 \vert 1)$ and its associated Lie group to describe its dynamics \cite{Buzano1989}.
A unitary rotation of the fermion mode on the Bloch sphere \cite{RodriguezLara2005} transforms the JC model into the anti-Jaynes--Cummings (aJC) model \cite{Solano2003}, 
\begin{align}
     \frac{\hat{H}}{\hbar} =&~ \omega \hat{a}^{\dagger} \hat{a} - \frac{1}{2} \omega_{0} \hat{\sigma}_{z}  - \lambda \left( \hat{a}^{\dagger} \hat{\sigma}_{+} + \hat{a} \hat{\sigma}_{-} \right).
\end{align}
These isomorphic models show energy level crossings, collapse-revival of the fermion population inversion \cite{Eberly1985,Rempe1987}, entanglement between radiation and matter \cite{Phoenix1988,GeaBanacloche1991}, and boson blockade \cite{Birnbaum2005}, to mention a few.

Building upon the JC and aJC models, the anisotropic Rabi (AR) model ~\cite{Xie2014}, 
\begin{align}
   \begin{aligned}
     \frac{\hat{H}}{\hbar} =&~ \omega \hat{a}^{\dagger} \hat{a} + \frac{1}{2} \omega_{0} \hat{\sigma}_{z}  + \lambda \left( \hat{a} \hat{\sigma}_{+} + \hat{a}^{\dagger} \hat{\sigma}_{-} \right)  + \mu \left( \hat{a}^{\dagger} \hat{\sigma}_{+} + \hat{a} \hat{\sigma}_{-} \right)+ \Omega \hat{\sigma}_{x},
   \end{aligned}
\end{align}
introduces an additional layer of complexity and generality beyond the quantum Rabi model \cite{Braak2016}.
The AR model is feasible of simulation in superconducting circuit-QED \cite{Wang2018},  trapped-ion-QED \cite{Wang2019}, and magnon-QED \cite{Skogvoll2021} where it may help providing quantum state engineering.  
The linear driving in the model breaks parity but helps in building exceptional solutions when the couplings are proportional to the boson frequency \cite{Li2021}. 
This exceptional regime is usually referred as a hidden symmetry.

The parity preserving AR model, without linear driving $\Omega = 0$, shows an underlying symmetry provided by the superalgebra $osp(2 \vert 2)$ ~\cite{Buzano1989, Schmitt1990} that includes two super-symmetric (SUSY) sectors~\cite{BerubeLauziere1994,Hussin2005}, related to JC and aJC dynamics, and a special unitary group sector, related to boson mode squeezing.
The spectra for the parity preserving AR model has been studied in exceptional \cite{Tomka2014} and approximated \cite{Zhang2015, Zhang2017} regimes.
The first attempts to employ an operational approach to the undriven model provided parametric oscillator driving terms that were neglected in adequate regimes \cite{Zhang2016} and helped identify a superradiant phase \cite{Shen2017} that led to identifying universal scaling and critical exponents in the anisotropic Dicke model \cite{Liu2017}.
Recent results, that do not neglect the parametric oscillator driving terms, identified degenerate states as well as crossings in the spectrum utilizing a Bogoliubov approach \cite{Chen2021}.
Further research, including interaction with an environment, described its spectral properties in the strong-coupling regime \cite{GutierrezJauregui2021}, and the survival of quantum correlations \cite{Ye2023,Lyu2023,Xu2024}.

From a mathematical physics perspective, identifying the underlying algebraic structure of a physical system is crucial for understanding its properties and dynamics.
However, translating those elegant mathematical constructs to tangible experimental observations may present significant challenges. 
Here, we are interested in bridging the gap between the abstract algebra structure behind these three models and its connection with the physics that may be observed in the laboratory, focusing on their proper states and spectra.
We aim to highlight the importance of an operator-based approach provided by Lie theory as well as the nuances required to bring its mathematical elegance into experimental observables.
In the following, we review in Sec. \ref{sec:Sec2} the SUSY structure of the JC model to provide its proper states and energies, as well as critical couplings providing energy crossing.
In Section \ref{sec:Sec3}, we review the aJC model which is isomorphic to the JC model up to an unitary rotation and, in consequence, has the same underlying structure. 
We investigate the SUSY structure of the AR model in Sec. \ref{sec:Sec4} and approximate its spectrum.
In Section \ref{sec:Sec5}, we focus on an exceptional regime isomorphic to the standard unbroken SUSY Hamiltonian that shows no spectral crossings.
We close with our conclusions in Sec. \ref{sec:Sec6}.

\section{Jaynes--Cummings model}
\label{sec:Sec2}
The Jaynes--Cummings (JC) model \cite{JaynesCummings1963},
\begin{align}
    \frac{\hat{H}_{JC}}{\hbar} = \omega \hat{a}^{\dagger} \hat{a} + \frac{1}{2}  \omega_0 \hat{\sigma}_z   + \lambda \left( e^{i \theta} \hat{a} \hat{\sigma}_{+} + e^{-i \theta} \hat{a}^{\dagger} \hat{\sigma}_{-} \right),
\label{Hjc}
\end{align}
describes the interaction of a boson field mode, characterized by frequency $\omega$ and annihilation (creation) operators $\hat{a}$ ($\hat{a}^{\dagger}$), and a spin-half fermion mode, characterized by frequency $\omega_0$ and lowering (raising) operators $\hat{\sigma}_{-}$ ($\hat{\sigma}_{+}$). 
The complex coupling strength $\lambda e^{i \theta}$ of the interaction models the amplitude and phase, $\lambda \ge 0$ and $0 \le \theta \le 2 \pi$, arising from the correlated annihilation (creation) of a boson mode excitation  with the creation (annihilation) of a fermion mode excitation.
The JC model belongs to a Hamiltonian class with an underlying SUSY provided by a graded Lie algebra \cite{MaldonadoVillamizar2021}.

The $D=2$ SUSY structure underlying the JC model is given in terms of the SUSY exchange operators \cite{Hussin2005} or super-bosons \cite{Stenholm2013},
\begin{align}
    \hat{Q}_{+} =  \hat{a} \hat{\sigma}_{+} \qquad \left( 
    \hat{Q}_{-} = \hat{a}^\dagger \hat{\sigma}_{-} \right),
    \label{eq:Qeop}
\end{align}
transforming excitations between boson and fermion modes.
The nilpotent exchange operators, $\hat{Q}_{\pm}^2 = 0$, help to define a SUSY Hamiltonian, 
\begin{align}
    \hat{H}_{Q} = \left\{ \hat{Q}_{+}, \hat{Q}_{-}\right\} = \hat{a}^{\dagger} \hat{a} + \dfrac{1}{2} \left( 1 + \hat{\sigma}_z \right) \equiv \hat{N}_{+}=\hat{N}_B+\hat{N}_F, \label{eq:HQ}
\end{align}
that is the total excitation number \cite{Ackerhalt1975}, in terms of the boson (fermion) excitation number operators $\hat{N}_B = \hat{a}^{\dagger} \hat{a}$ ($\hat{N}_F =  \left( 1 + \hat{\sigma}_z \right)/2 $).
The SUSY Hamiltonian serves as the Casimir invariant, $ \left[ \hat{H}_{Q}, \hat{Q}_{\pm} \right] = 0$ , of the superalgebra isomorphic to $u(1 \vert 1)$ \cite{Buzano1989}.
It decomposes into boson (fermion) sector,
\begin{align}
        \hat{H}_{B} = \hat{Q}_- \hat{Q}_{+}
        \qquad 
        \left(\hat{\hat{H}}_{F} = \hat{Q}_{+}\hat{Q}_- \right),
\end{align}
that the exchange operators intertwine, $ \hat{Q}_- \hat{H}_{F} = \hat{H}_{B}  \hat{Q}_- $ or $ \hat{H}_{F} \hat{Q}_{+} = \hat{Q}_{+} \hat{H}_{B}$, with the spectrum,
\begin{align}
     \hat{H}_{F} \vert e, n \rangle = (n+1) \vert e,n\rangle, \qquad \left( \hat{H}_{B} \vert g, n \rangle = n \vert g,n \rangle \right),
\end{align}
provided by the product of the excited (ground) state of the fermion mode and the number states of the boson mode.
Unbroken SUSY is evident as the ground state of the SUSY Hamiltonian is unique, has zero energy, and corresponds to the boson sector ground state.
This will play a fundamental role in diagonalizing the JC model using a Lie theory approach.

The adjoint nature of the exchange operators leads to the construction of two Hermitian SUSY charge operators, 
\begin{align}
    \hat{Q}_x = \hat{Q}_{+} + \hat{Q}_-, \qquad
    \hat{Q}_y = - i \left( \hat{Q}_{+} - \hat{Q}_- \right),
    \label{eq:charges}
\end{align}
whose square yields the SUSY Hamiltonian, 
\begin{align}
     \hat{Q}_x^2 = \hat{Q}_y^2 = \frac{1}{2} \left( \hat{Q}_x^2 + \hat{Q}_y^2 \right) = \hat{H}_{Q} = \hat{N}_{+},
     \label{eq:hqn+}
\end{align}
confirming a double degenerate spectrum, except for the zero-energy ground state.

In the SUSY framework, the JC model takes the form, 
\begin{align}
    \frac{\hat{H}_{JC}}{\hbar} = \omega \hat{N}_{+} + \Delta \hat{S}_z + \lambda \left( e^{i \theta} \hat{Q}_{+} + e^{-i \theta} \hat{Q}_-  \right) - \frac{\omega}{2},
\end{align}
with the detuning $\Delta = \omega_0 - \omega$, and the dimensionless fermion mode $z$-spin operator,
\begin{align}
 \hat{S}_z = \dfrac{1}{2} \hat{\sigma}_z, \qquad \left[ \hat{S}_z,   \hat{Q}_{\pm} \right] = \pm \hat{Q}_{\pm}, \qquad
  \left[ \hat{Q}_{+},   \hat{Q}_- \right] = 2 \hat{H}_{Q} \hat{S}_z,
\end{align}
providing a \textit{deformed} $su(2)$ algebra sector.
Moving to a reference frame defined by the total excitation number and a fixed rotation around the boson mode excitation number, 
\begin{align}
\vert \psi_{JC} \rangle = e^{-i \theta \hat{N}_B }e^{-i \omega (\hat{N}_{+} - 1/2) t } \vert \psi_{S} \rangle,
\end{align}
yields a Hamiltonian in the \textit{deformed} $su(2)$ sector,
\begin{align}
    \frac{\hat{H}_{S}}{\hbar} = \Delta \hat{S}_z + \lambda  \hat{Q}_x.
    \label{eq:HS}
\end{align}
At this point, we could follow standard SUSY procedure \cite{Hussin2005, MaldonadoVillamizar2021}.
However, it is possible to further simplify the approach by transitioning from the \textit{deformed} $su(2)$ to the standard $su(2)$ algebra.

The SUSY Hamiltonian zero-energy state is also an eigenstate of the JC Hamiltonian,
\begin{align}
\hat{H}_{Q}  \vert g, 0 \rangle = 0  \vert g, 0 \rangle,  \qquad \hat{H}_{JC} \vert g, 0 \rangle  = - \frac{\hbar}{2} \omega_{0} \vert g, 0 \rangle.
\end{align}
This suggests parceling the Hilbert space, $\mathcal{H} = \oplus \mathcal{H}_{j}^{(JC)}$, into orthogonal invariant subspaces~\cite{Hussin2005},
\begin{align}
     \quad \mathcal{H}_{0}^{(JC)}= \{ \vert g, 0 \rangle \}, \quad \mathcal{H}_{N}^{(JC)} = \{ \vert g, N \rangle, \vert e, N-1 \rangle \},
\end{align}
where we can identify the unique ground state subspace $\mathcal{H}_{0}^{(JC)}$ and the excited subspaces $\mathcal{H}_{N}^{(JC)}$ with constant excitation number $N = 1, 2, 3, \ldots$.
In the latter, the SUSY Hamiltonian is positive definite, $\langle \hat{\mathcal{H}} \rangle = \langle \hat{N}_{+} \rangle = N \ge 1$, invertible, and possesses a square root. 
Working in the excited subspaces allows the SUSY Hamiltonian to be replaced by the total excitation number operator,
\begin{align}
    \begin{aligned}
         &~ \hat{N}_{+}^{-1/2} \hat{Q}_{x} \left\vert \xi, N - \delta_{e,\chi} \right\rangle = \hat{Q}_{x} \hat{N}_{+}^{-1/2} \left\vert \chi, N - \delta_{e,\chi} \right\rangle ,  \\
         &~ \hat{N}_{+}^{-1/2} \hat{Q}_{y} \left\vert \chi, N - \delta_{e,\chi} \right\rangle = \hat{Q}_{y} \hat{N}_{+}^{-1/2} \left\vert \chi, N - \delta_{e,\chi} \right\rangle,
    \end{aligned}
\end{align}
with $\chi=g,e$ and the Kronecker delta $\delta_{i,j}$.
This helps us to define a $su(2)$ algebra for the {\em excited} subspaces,
\begin{align}
    \hat{S}_{x}^{(Q)} = \frac{1}{2} \hat{N}_{+}^{-1/2} \hat{Q}_{x}, \qquad
    \hat{S}_{y}^{(Q)} = \frac{1}{2} \hat{N}_{+}^{-1/2} \hat{Q}_{y}, 
\end{align}
whereby the \textit{deformed} $su(2)$ Hamiltonian becomes the standard Rabi Hamiltonian,
\begin{align}
    \frac{\hat{H}_{S}^{(Q)}}{\hbar} = \Delta \hat{S}_z + \hat{\Lambda}(\hat{N}_{+}) \hat{S}^{(Q)}_{x}.
\end{align}
The coupling operator, 
\begin{align}
    \hat{\Lambda}(\hat{N}_{+}) =  2 \lambda \hat{N}_{+}^{1/2},
\end{align}
enhances the boson-fermion coupling strength by a factor proportional to the square root of the total excitation number, which is an invariant of the algebra. 
A unitary rotation on the Bloch sphere, 
\begin{align}
\begin{aligned}
\vert \psi_{S}^{(Q)} \rangle =&~ e^{-i \hat{\beta}(\hat{N}_{+}) \hat{S}^{(Q)}_{y}} \vert \psi_{D}^{(Q)} \rangle,
\end{aligned}
 \quad  \tan \hat{\beta}(\hat{N}_{+}) = \frac{ \hat{\Lambda}(\hat{N}_{+}) }{\Delta} ,
\end{align}
with the rotation angle $\langle \beta \rangle \in [-\pi/2, \pi/2]$ for the model parameter ranges, diagonalizes the Hamiltonian, 
\begin{align}
    \frac{\hat{H}_{D}^{(Q)}}{\hbar} = \sqrt{\Delta^{2} + 4 \lambda^{2} \hat{N}_{+}}\, \hat{S}_{z}, 
    \label{eq:HD}
\end{align}
to that of a free fermion mode with an effective Rabi frequency operator,
\begin{align}
   \hat{\Omega}(\hat{N}_{+}) = \sqrt{\Delta^{2} + 4 \lambda^{2} \hat{N}_{+}}, 
\end{align}
proportional to the total excitation number.

The energy spectrum for the full Hilbert space,
\begin{align}
    \begin{aligned}
        &\vert g , N \rangle ,  &E_{g,N}^{(Q)} =  - \frac{\hbar}{2} \sqrt{\Delta^2 + 4 \lambda^2 N}, \\
        &\vert e , N-1 \rangle ,  &E_{g,N-1}^{(Q)} = \frac{\hbar}{2} \sqrt{\Delta^2 + 4 \lambda^2 N}, 
    \end{aligned}
\end{align}
with $N=1,2,\ldots$, helps us to realize the squared JC model, 
\begin{align}
    \left( \frac{\hat{H}_{D}^{(Q)}}{\hbar} \right)^{2} = \frac{1}{4} \left( \Delta^2 + 4 \lambda^2 \hat{N} \right),
\end{align}
having the structure of non-interacting free fermion and boson modes with unbroken SUSY. 
In particular, the ground state is unique and belongs to the boson sector, while the rest of the spectrum is double degenerate. 
The nearest-neighbor energy gap is constant, $E_{\chi,N+1}^{(Q) 2} - E_{\chi,N}^{(Q) 2} = \lambda^2$, and proportional to the squared coupling strength.

The eigenstates of the JC model in excited subspaces in the original frame are the Gilmore-Perelomov coherent states, 
\begin{align}
    \begin{aligned}
    \vert {\scriptstyle{\pm}}; N \rangle_{JC} =&~ e^{-i \theta \hat{N}_B} e^{- i \omega \hat{N}_{+} t}  e^{-i \hat{\beta}(\hat{N}_{+}) \hat{S}^{(Q)}_{y}} \vert \chi,n \rangle,
    \end{aligned}
\end{align}
generated by the action of a homomorphism of the $su(2)$ sector on the diagonal frame eigenstates, where the positive (negative) label corresponds to the ground (excited) state.
The full set of eigenstates, 
\begin{align}
 \begin{aligned}
    \vert -; 0 \rangle_{JC} =&~ \vert g, 0 \rangle, \\
    \vert {\scriptstyle\pm}; N \rangle_{JC} =&~ \frac{1}{\sqrt{2}} \left[ \sqrt{ \frac{\Omega(N) + \Delta}{\Omega(N)}} \left\vert \chi, N - \delta_{{\scriptscriptstyle +},{\scriptscriptstyle{\pm}}} \right\rangle + 
    \pm  \sqrt{ \frac{ \Omega(N) - \Delta}{\Omega(N)}} \vert \tilde{\chi}, N -\delta_{{\scriptscriptstyle -},{\scriptscriptstyle \pm}} \rangle \right], 
 \end{aligned}
\end{align}
where $\chi\in \{ g, e\}$, $\tilde\chi\in \{\tilde g, \tilde e$\}, $\tilde{g} = e$, $\tilde{e} = g$, is given in terms of the effective Rabi frequency,
\begin{align}
    \Omega(N) = \sqrt{\Delta^2 + 4 \lambda^2 N},
\end{align}
for $N=1,2,3, \ldots$ 
The excited subspaces provide pairs of the Bell type states that coincide with the standard Bell states for $N=1$ \cite{Phoenix1988,GeaBanacloche1991}.
Their fermion and boson modes reduced density matrices, 
\begin{align}
    &\begin{aligned}
        \rho_{\pm,N}^{(JC,F)} =& \cos^2 \frac{\beta(N)}{2} \vert x \rangle \langle x \vert + \sin^2 \frac{\beta(N)}{2} \vert \tilde{x} \rangle \langle \tilde{x} \vert \\
        =& \frac{1}{2 \Omega(N)} \left\{ \left[ \Omega(N) + \Delta \right] \vert x \rangle \langle x \vert +  \left[ \Omega(N) - \Delta \right] \vert \tilde{x} \rangle \langle \tilde{x} \vert \right\},
    \end{aligned} \\
    &\begin{aligned}
        \rho_{\pm,N}^{(JC,B)} =&~ \cos^2 \frac{\beta(N)}{2} \left\vert N -\delta_{-,\mp}  \right\rangle \left\langle N - \delta_{-,\mp} \right\vert +  \sin^2 \frac{\beta(N)}{2} \left\vert N - \delta_{+,\mp} \right\rangle \left\langle N -\delta_{+,\mp} \right\vert \\
        =& \frac{1}{2}  \left\{ \frac{ \Omega(N) + \Delta }{\Omega(N)}\left\vert N -\delta_{-,\mp}  \right\rangle \left\langle N - \delta_{-,\mp} \right\vert +  \frac{\Omega(N) - \Delta }{\Omega(N)} \left\vert N -\delta_{+,\mp} \right\rangle \left\langle N - \delta_{+,\mp} \right\vert \right\},
    \end{aligned}
\end{align}
confirm that maximal entanglement occurs at resonance $\Delta = 0$.

\begin{figure}[!ht]
\centering
\fbox{\includegraphics[scale=1]{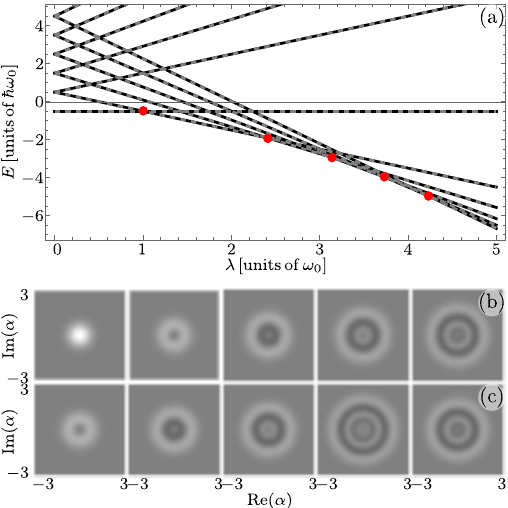}}
\caption{(a) Eleven energy levels of the JC model spectrum, $\left\{ \vert -, 0 \rangle_{JC}, \vert \pm, N \rangle_{JC} \right\}$, on resonance $\Delta = 0$.
Wigner quasi-probability distribution for (b) the ground state and (c) the first excited state at the critical coupling $\lambda_{N}$ for $N=1, 2, 3, 4, 5$ marked with circles in (a).}
\label{fig:Fig1}
\end{figure}
The energy spectrum in the original frame,
\begin{align}
 \begin{aligned}
    E_{-,0}^{(JC)} =&~ - \frac{\hbar}{2} \omega_0 , \\
    E_{\pm,N}^{(JC)} =&~ \hbar \omega \left( N - \frac{1}{2} \right) \pm \hbar\sqrt{ \Delta^2 + 4 \lambda^2 N}, 
 \end{aligned}
\end{align}
allows calculating all possible crossings at critical couplings,
\begin{align}
    \lambda_{\pm M, -N} = \sqrt{ \omega \left[ \left( M + N \right) \omega \mp \sqrt{  \omega_0^2 - 2 \omega_0 \omega + (4 M N + 1) \omega^2 } \right]},
\end{align}
for $N>M$ such that $E_{\pm,M}\vert_{\lambda_{\pm M, - N}} = E_{-,N}\vert_{\lambda_{\pm M, -N}}$.
In particular, the state $\vert -, N \rangle$ becomes the ground state just after the critical couplings, 
\begin{align}
 \lambda_{N} = \sqrt{ \omega \left[ \left( 2N -1 \right) \omega + \sqrt{  \omega_0^2 - 2 \omega_0 \omega + (2N - 1)^2 \omega^2 } \right]}.
\end{align}
At each of these critical couplings the lowest energy is degenerate. 
Note that, for a boson mode frequency tuned to the red of the fermion mode frequency, $\omega_0 > \omega$, $\lambda_{1}$ reduces to the well known first crossing $\lambda_{1} = \sqrt{ \omega \omega_0}$.

Figure \ref{fig:Fig1}(a) shows eleven elements of the JC model spectrum. Fig. \ref{fig:Fig1}(b) shows the Wigner function for the boson mode reduced density matrix \cite{Cahill1969}, 
\begin{align}
    \begin{aligned}
        W_{-, 0}(\alpha) =&~ \frac{1}{2 \pi}  e^{- 2 \vert \alpha \vert^2}, \\
        W_{\pm, N}(\alpha) =&~ \frac{(-1)^{N} e^{- 2 \vert \alpha \vert^2} }{\pi \Omega(N)} \left\{ \left[ \Omega(N) \mp \Delta \right] L_{N} (4 \vert \alpha \vert^2) - \left[ \Omega(N) \pm \Delta \right] L_{N-1} (4 \vert \alpha \vert^2) \right\},
    \end{aligned}
\end{align}
at the critical couplings on resonance, $\lambda_{N} = (  \sqrt{N} + N - 1 )  \omega_0$, where $L_{n}(x)$ are the Laguerre polynomials.
Fig. \ref{fig:Fig2}(c) displays the first excited state at that critical crossing.

\section{anti-Jaynes--Cummings model}
\label{sec:Sec3}
The anti-Jaynes--Cummings (aJC) model,
\begin{align}
    \frac{\hat{H}_{aJC}}{\hbar} = \omega \hat{a}^{\dagger} \hat{a} - \dfrac{1}{2} \omega_0  \hat{\sigma}_z  - \lambda \left( e^{-i \theta} \hat{a}^{\dagger} \hat{\sigma}_{+} + e^{i \theta} \hat{a} \hat{\sigma}_-  \right),
\end{align}
is isomorphic to the JC model under a $(\pi/2)$-rotation around 
the Pauli-$y$ operator \cite{RodriguezLara2005},
\begin{align}
    \vert \psi_{JC} \rangle = e^{- i \frac{\pi}{2} \hat{\sigma}_y } \vert \psi_{aJC} \rangle,
\end{align}
providing the so-called counter-rotating terms, where an excitation in both fermion and boson modes is either created or destroyed.
All results of Sec. \ref{sec:Sec2} for the JC model hold under this unitary rotation.
For the sake of completeness we show them explicitly.

The structure of the aJC is provided by alternative SUSY exchange operators of the $osp(2 \vert 2)$ superalgebra~\cite{Buzano1989, Schmitt1990}, 
\begin{align}
    \hat{R}_- =  \hat{a}^{\dagger} \hat{\sigma}_{+} \qquad \left( 
    \hat{R}_{+} = \hat{a} \hat{\sigma}_- \right),
    \label{eq:Reop}
\end{align}
that create (annihilate) an excitation in both the fermion and boson modes, Eq. (\ref{eq:Qeop}).
The corresponding SUSY Hamiltonian, Eq. (\ref{eq:HQ}),
\begin{align}
    \hat{H}_{R} = \left\{ \hat{R}_{+}, \hat{R}_- \right\} = \hat{a}^{\dagger} \hat{a} + \dfrac{1}{2} \left( 1 - \hat{\sigma}_z \right) \equiv \hat{N}_-,
\end{align}
plays the role of the total excitation number operator in the new frame.
The Hermitian SUSY charges (cf. Eq. (\ref{eq:charges})), 
\begin{align}
    \begin{aligned}
        \hat{R}_{x} = \hat{R}_{+} + \hat{R}_-, \qquad 
        \hat{R}_{y} = - i \left( \hat{R}_{+} - \hat{R}_- \right),
    \end{aligned}
\end{align}
yield the SUSY Hamiltonian (cf. Eq. (\ref{eq:hqn+})), 
\begin{align}
    \hat{R}_x^2 = \hat{R}_y^2 = \frac{1}{2} \left( \hat{R}_x^2 + \hat{R}_y^2 \right) = \hat{H}_{R},
\end{align}
that is the Casimir invariant of the super-algebra, $\left[ \hat{H}_{R}, \hat{R}_{j} \right]=0$, $j= \pm, x, y$. 

The SUSY Hamiltonian zero-energy state is an eigenstate of the aJC Hamiltonian,
\begin{align}
\hat{H}_{R}  \vert e, 0 \rangle = 0  \vert e, 0 \rangle,  \qquad \hat{H}_{aJC} \vert e, 0 \rangle  = - \frac{\hbar}{2} \omega_{0} \vert e, 0 \rangle,
\end{align}
suggesting parceling the Hilbert space, $\mathcal{H} = \oplus \mathcal{H}_{j}^{(aJC)}$,
\begin{align}
    \mathcal{H}_{0}^{(aJC)} = \{ \vert e, 0 \rangle \}, \quad \mathcal{H}_{N}^{(aJC)} = \{ \vert e, N \rangle, \vert g, N-1 \rangle \},
 \end{align}
into the ground state subspace and constant excitation number subspaces with $N~\text{=}~1,~2,~3,\ldots$, where the SUSY Hamiltonian is positive definite, $\langle \hat{H}_{R} \rangle = \langle \hat{N}_- \rangle = N \ge 1$, invertible, possesses a square root, and commutes with the SUSY charge operators.
In full analogy to Sec. \ref{sec:Sec2}, working in the excited subspaces allows replacing the SUSY Hamiltonian by the corresponding total excitation number operator, 
\begin{align}
    \begin{aligned}
        &\hat{N}_{-}^{-1/2} \hat{R}_{x} \left\vert \chi, N - \delta_{g,\chi} \right\rangle = \hat{R}_{x} \hat{N}_{-}^{-1/2} \left\vert \chi, N - \delta_{g,\chi} \right\rangle, \\
        & \hat{N}_{-}^{-1/2} \hat{R}_{y} \left\vert \chi, N - \delta_{g,\chi} \right\rangle = \hat{R}_{y} \hat{N}_{-}^{-1/2} \left\vert \chi, N - \delta_{g,\chi} \right\rangle ,       
    \end{aligned}
\end{align}
with $\chi=g,e$.
The corresponding $su(2)$ algebra sector for the excited subspaces,
\begin{align}
    \hat{S}_{x}^{(R)} = \frac{1}{2} \hat{N}_{-}^{-1/2} \hat{R}_{x}, \qquad
    \hat{S}_{y}^{(R)} = \frac{1}{2} \hat{N}_{-}^{-1/2} \hat{R}_{y}, 
\end{align}
yields a Rabi Hamiltonian, 
\begin{align}
    \frac{\hat{H}_{S}^{(R)}}{\hbar} = -\Delta \hat{S}_z - \hat{\Lambda}(\hat{N}_{-}) \hat{S}^{(R)}_x,
\end{align}
with coupling operator,
\begin{align}
    \hat{\Lambda}(\hat{N}_{-}) = 2 \lambda \hat{N}_{-} ^{1/2}.
\end{align}
One arrives at the Rabi Hamiltonian by moving into the reference frame,
\begin{align}
\vert \psi_{aJC} \rangle = e^{-i \theta \hat{N}_B } e^{-i \omega (\hat{N}_{-} - 1/2) t } \vert \psi_{S}^{(R)} \rangle,
\end{align}
provided by the boson excitation number and the total excitation number in this reference frame.
A rotation on the Bloch sphere, 
\begin{align}
\begin{aligned}
\vert \psi_{S}^{(R)} \rangle =&~ e^{i \hat{\beta}(\hat{N}_{-}) \hat{S}^{(R)}_{y}} \vert \psi_{D}^{(R)} \rangle,
\end{aligned}
 \quad  \tan \hat{\beta}(\hat{N}_{-}) = \frac{ \hat{\Lambda}(\hat{N}_{-}) }{\Delta} ,
\end{align}
with rotation angle $\langle \beta \rangle \in [0, \pi/2]$, diagonalizes the model, 
\begin{align}
    \frac{\hat{H}_{D}^{(R)}}{\hbar} = -\sqrt{\Delta^{2} + 4 \lambda^{2} \hat{N}_{-}} \, \hat{S}_z,
\end{align}
to that of a free fermion mode with effective Rabi frequency proportional to the total excitation number with energy spectrum,
\begin{align}
    \begin{aligned}
        &\vert e , N \rangle ,  &E_{e,N}^{(R)} =  - \frac{\hbar}{2} \sqrt{\Delta^2 + 4 \lambda^2 N}, \\
        &\vert g , N-1 \rangle ,  &E_{g,N-1}^{(R)} = \frac{\hbar}{2} \sqrt{\Delta^2 + 4 \lambda^2 N}, 
    \end{aligned}
\end{align}
with $N=1,2,\ldots$ that helps us realize that the squared aJC model in the diagonal reference frame for the full Hilbert space, 
\begin{align}
    \left(\frac{\hat{H}_{D}^{(R)}}{\hbar}\right)^{2} = \frac{1}{4} \left[ \Delta^2 + 4 \lambda^2 \hat{N}_{-} \right],
\end{align}
has an unbroken SUSY structure; that is, a  unique ground state, while the rest of the spectrum is double degenerate with constant nearest-neighbor energy gap, $E_{\chi,N+1}^{(R)2}~\text{-}~E_{\chi,N}^{(R)2}~=~\lambda^2$, proportional to the squared coupling strength.

The full set of eigenstates of the aJC model,
\begin{align}
 \begin{aligned}
    \vert -; 0 \rangle_{aJC} =&~ \vert e, 0 \rangle, \\
    \vert \pm; N \rangle_{aJC} =&~ \frac{1}{\sqrt{2}} \left[ \sqrt{ \frac{\Omega(N) + \Delta}{\Omega(N)}} \left\vert x, N - \delta_{-,\pm} \right\rangle \pm  \sqrt{ \frac{ \Omega(N) - \Delta}{\Omega(N)}} \vert \tilde{x}, N -\delta_{+,\pm} \rangle \right], 
 \end{aligned}
\end{align}
has energy spectrum,
\begin{align}
 \begin{aligned}
    E_{-,0}^{(aJC)} =&~ - \frac{\hbar}{2} \omega_0 , \\
    E_{\pm,N}^{(aJC)} =&~ \hbar \omega \left( N - \frac{1}{2} \right) \pm \hbar \sqrt{ \Delta^2 + 4 \lambda^2 N}, 
 \end{aligned}
\end{align}
identical to the JC model spectrum.
Consequently, the energy crossings occur at the same critical couplings.

\begin{figure}[!ht]
\centering
\fbox{\includegraphics[scale=1]{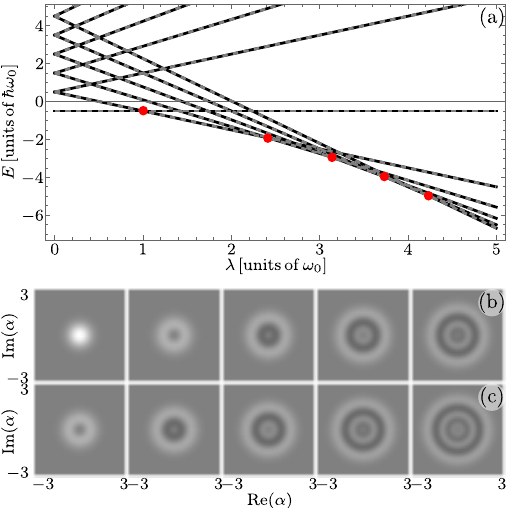}}
\caption{Same as Fig. \ref{fig:Fig1} but for the aJC model.}
\label{fig:Fig2}
\end{figure}
The excited subspaces eigenvectors in the original frame, 
\begin{align}
    \begin{aligned}
    \vert \pm; N \rangle_{aJC} =&~ e^{-i \theta \hat{N}_B} e^{- i \omega \hat{N}_{-} t}  e^{i \hat{\beta}(\hat{N}_{-}) \hat{S}^{(R)}_{y}} \vert x,n \rangle ,
    \end{aligned}
\end{align}
are the Gilmore-Perelomov coherent states with reduced density matrices, 
\begin{align}
    &\begin{aligned}
        \rho_{\pm,N}^{(aJC,F)} =& \cos^2 \frac{\beta(N)}{2} \vert \tilde{\chi} \rangle \langle \tilde{\chi} \vert + \sin^2 \frac{\beta(N)}{2} \vert \chi \rangle \langle \chi \vert \\
        =& \frac{1}{2 \Omega(N)} \left\{ \left[ \Omega(N) + \Delta \right] \vert \tilde{\chi} \rangle \langle \tilde{\chi} \vert + \left[ \Omega(N) - \Delta \right] \vert \chi \rangle \langle \chi \vert \right\},
    \end{aligned} \\
    &\begin{aligned}
        \rho_{\pm,N}^{(aJC,B)} =&~ \cos^2 \frac{\beta(N)}{2} \left\vert N -\delta_{-,\pm}  \right\rangle \left\langle N - \delta_{-,\pm} \right\vert +   \sin^2 \frac{\beta(N)}{2} \left\vert N - \delta_{+,\pm} \right\rangle \left\langle N -\delta_{+,\pm} \right\vert \\
        =& \frac{1}{2}  \left\{ \frac{ \Omega(N) + \Delta }{\Omega(N)}\left\vert N -\delta_{-,\pm}  \right\rangle \left\langle N - \delta_{-,\pm} \right\vert +  \frac{\Omega(N) - \Delta }{\Omega(N)} \left\vert N -\delta_{+,\pm} \right\rangle \left\langle N - \delta_{+,\pm} \right\vert \right\},
    \end{aligned}
\end{align}
that provide the Bell type states.

Figure \ref{fig:Fig2}(a) shows eleven elements of the aJC model spectrum. Fig. \ref{fig:Fig2}(b) shows the Wigner function for the boson mode reduced density matrix.

\section{Anisotropic Rabi model}
\label{sec:Sec4}
We are interested in a parity preserving version of the anisotropic quantum Rabi model, 
\begin{align}
\begin{aligned}
    \frac{\hat{H}_{aR}}{\hbar} =&~ \omega \hat{a}^{\dagger} \hat{a} + \frac{1}{2} \omega_0 \hat{\sigma}_z + \lambda \left( e^{i \theta} \hat{a} \hat{\sigma}_{+} + e^{-i \theta} \hat{a}^{\dagger} \hat{\sigma}_{-} \right)   + \mu \left( e^{-i \theta} \hat{a}^{\dagger} \hat{\sigma}_{+} + e^{i \theta} \hat{a} \hat{\sigma}_{-} \right), 
\end{aligned}
\end{align}
where the aJC coupling strength $\mu$ is real positive, $\mu \ge 0$, and cannot take values equal to the JC coupling strength, $\mu \neq \lambda$. 
In the SUSY framework, 
\begin{align}
    \begin{aligned}
       \frac{\hat{H}_{aR}}{\hbar} = &~\omega \hat{N}_{B} + \omega_0 \hat{S}_z + \lambda \left( e^{i \theta} \hat{Q}_{+} + e^{-i \theta} \hat{Q}_{-}  \right) +  \mu \left( e^{i \theta} \hat{R}_{+} + e^{-i \theta} \hat{R}_{-}  \right),
    \end{aligned}
\end{align}
it involves the two SUSY sectors discused before that unveil, 
\begin{align}
    \left\{ \hat{Q}_{\pm}, \hat{R}_{\pm} \right\} = 2 \hat{K}_{\mp},
\end{align}
a $su(1,1)$ algebra sector,
\begin{align}
    \left[ \hat{K}_z, \hat{K}_{\pm} \right] = \pm \hat{K}_{\pm}, \qquad
    \left[ \hat{K}_{+}, \hat{K}_{-} \right] = - 2 \hat{K}_z,
\end{align}
with two representations characterized by Bargmann parameters $k = 1/4, 3/4$ for even and odd boson excitation states, in that order.
We define the operator $\hat{K}_z = (\hat{N}_B + 1/2 )/2$ and the Casimir operator $\hat{K}^2 = \hat{K}_{z}^{2} - \{ \hat{K}_{+}, \hat{K}_{-} \}/2$, i.e. the invariant of the algebra, taking the value $\hat{K}^{2} = -3/16$ in the even and odd boson excitation subspaces.
Special attention is required for the single boson mode representation of the $su(1,1)$ algebra \cite{ArmentaRico2020,Onah2023}, where the Hermitian charge operators,
\begin{align}
   \begin{aligned}
      \hat{K}_x =&~ \frac{1}{4} \left( \hat{a}^2 + \hat{a}^{\dagger 2} \right) = - \frac{1}{4} \left( \hat{p}^2 - \hat{q}^2 \right), \\    
      \hat{K}_y =&~ -\frac{i}{4} \left( \hat{a}^{\dagger 2}  - \hat{a}^2\right) = - \frac{1}{4} \left\{ \hat{q},  \hat{p} \right\}, \\
      \hat{K}_z =&~ \frac{1}{4} \left\{ \hat{a}^{\dagger},  \hat{a} \right\} =  \frac{1}{4} \left( \hat{p}^2 + \hat{q}^2 \right),
   \end{aligned}
\end{align}
involve the inverted quantum harmonic oscillator $\hat{K}_x$.
We used the dimensionless canonical pair $\hat{q} = \left( \hat{a}^{\dagger} + \hat{a} \right) / \sqrt{2}$ and $\hat{p} = i \left( \hat{a}^{\dagger} - \hat{a} \right) / \sqrt{2}$.
The Hermitian charge operators $\hat{K}_{x}$ and $\hat{K}_{y}$ possess continuous spectrum; this is a noncompact algebra.

The existence of the $su(1,1)$ sector suggests employing a squeezing transformation \cite{Schmitt1990, GutierrezJauregui2021},
\begin{align}
\vert \psi_{aR} \rangle = e^{- \frac{i}{2} \omega t} e^{- i \theta \hat{N}_B}  e^{-i \Theta(\mu -\lambda) \frac{\pi}{2} \hat{\sigma}_{y}} e^{-i 2 \ln \frac{ \lambda + \mu }{|\lambda - \mu|} \hat{K}_y}   \vert \psi_{S} \rangle,
\end{align}
where we used Heaviside theta function $\Theta(x)$ to obtain an effective Hamiltonian, 
\begin{align}
    \begin{aligned}
        \frac{\hat{H}_{S}}{\hbar} =&~ \frac{\omega}{\vert \lambda^{2} - \mu^{2} \vert} \left[ 2 \left( \lambda^{2} + \mu^{2} \right) \hat{K}_{z} - 4 \lambda \mu \hat{K}_{x} \right] + \mathrm{sgn}(\lambda - \mu) \left[ \omega_{0} \hat{S}_{z} + \sqrt{\vert \lambda^{2} - \mu^{2} \vert} \hat{Q}_{x} \right],
    \end{aligned} 
\end{align}
describing a JC model with a parametric oscillator drive term.
Here the sign function has been defined as $\mathrm{sgn}(x) = x / \vert x \vert$ for $x \neq 0$.
This formulation highlights the isotropic Rabi model as a singular limit.
The latter is evidenced by the the divergence of the squeezing parameter, $ \xi = \ln ( \lambda + \mu )/(\vert\lambda - \mu\vert)$, 
for equal coupling strengths $\lambda=\mu$.
The squeezed frame effective model recovers the JC and aJC models, 
\begin{align}
    \begin{aligned}
        \lim_{\lambda \gg \mu} \hat{H}_{S} \approx &~ \hat{H}_{JC} , \\
        \lim_{\lambda \ll \mu} \hat{H}_{S} \approx &~   \left. \hat{H}_{JC} \right\vert_{\omega_{0} \rightarrow -\omega_{0}, \lambda \rightarrow -\mu } = \hat{H}_{aJC},
    \end{aligned}
\end{align}
in their respective limits.
The effective Hamiltonian, 
\begin{align}
    \begin{aligned}
        \frac{\hat{H}_{S}}{\hbar} =&~ \frac{\omega}{2} \left[ \frac{\lambda + \mu}{ \vert \lambda - \mu \vert} \hat{p}^{2} + \frac{\vert \lambda - \mu \vert}{ \lambda + \mu } \hat{q}^{2} \right]  + \mathrm{sgn}(\lambda - \mu)  \left[ \omega_0  \hat{S}_z + \sqrt{\vert \lambda^{2} - \mu^{2} \vert} ~ \hat{Q}_x \right],
    \end{aligned}
\end{align}
in terms of the dimensionless canonical pair suggests that the AR model does not show an equivalent spectral collapse to that of the Rabi model \cite{MaldonadoVillamizar2019}.

Assuming that the coupling strengths fulfill, 
\begin{align}
    2 \lambda \mu \ll  \lambda^{2} + \mu^{2},
\end{align}
let us approximate the effective Hamiltonian by neglecting the parametric oscillator drive term \cite{Zhang2016,Shen2017,Liu2017},
\begin{align}
    \begin{aligned}
        \frac{\hat{H}_{S}}{\hbar} \approx &~ 2  \frac{\lambda^{2} + \mu^{2}}{\vert \lambda^{2} - \mu^{2} \vert} \omega  \hat{K}_{z}  +\mathrm{sgn}(\lambda - \mu) \left[ \omega_{0} \hat{S}_{z} + \sqrt{\vert \lambda^{2} - \mu^{2} \vert} \hat{Q}_{x} \right].
    \end{aligned}
\end{align}
The approximation describes the JC model with a scaled boson frequency and a sign defined by the dominant coupling strength.
Hereafter, we follow our approach and split the corresponding Hilbert space into ground and excited subspaces to diagonalize the approximated effective Hamiltonian using a rotation on the Bloch sphere,
\begin{align}
    \vert \psi_{S} \rangle = e^{-i \frac{\lambda^2+ \mu^{2}}{\vert \lambda^2 - \mu^{2} \vert} \omega \hat{N}_{+} t} e^{- i \alpha(\hat{N}_{+}) \hat{S}_{y}^{(Q)} } \vert \psi_{DJC} \rangle,
\end{align}
with a rotation angle operator, 
\begin{align}
    \tan \alpha(\hat{N}_{+}) = \frac{2 \Lambda_{aR}}{\Delta_{aR}} \hat{N}_{+}^{1/2},
\end{align}
in terms of the effective detuning and coupling strength,
\begin{align}
   \begin{aligned}
     \Delta_{aR} =&~ \mathrm{sgn}(\lambda - \mu) \omega_{0} - \frac{\lambda^2+ \mu^{2}}{\vert \lambda^2 - \mu^{2} \vert} \omega,\\
     \Lambda_{aR}=&~  \mathrm{sgn}(\lambda - \mu) \sqrt{ \vert \lambda^2 - \mu^{2} \vert},
   \end{aligned}
\end{align}
in that order, that yield a diagonal effective Hamiltonian,
\begin{align}
   \begin{aligned}
    \frac{\hat{H}_{DJC}}{\hbar} =&~ \mathrm{sgn}(\lambda - \mu) \sqrt{ \Delta_{aR}^{2} + 4 \Lambda_{aR}^{2} \hat{N}_{+}} \, \hat{S}_{z}.
   \end{aligned}
\end{align}
The latter allows one to approximate the eigenstates, energy spectrum and critical couplings of the effective Hamiltonian in this regime in closed form that is cumbersome to write here.

\begin{figure}[!ht]
\centering
\fbox{\includegraphics[scale=1]{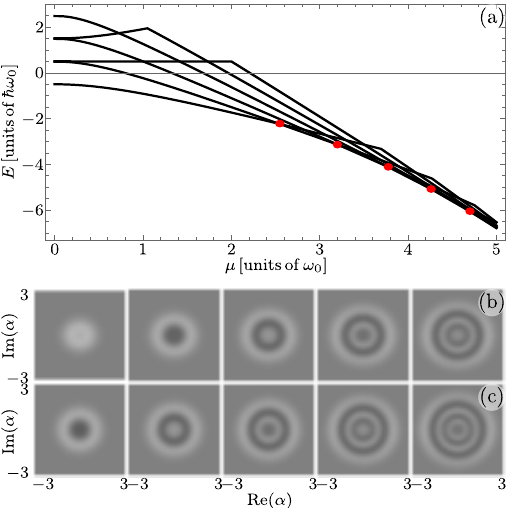}}
\caption{Same as Fig. \ref{fig:Fig1} but with the ARM for fixed $\lambda = 0.01 \omega_{0}$.}
\label{fig:Fig3}
\end{figure}
Figure \ref{fig:Fig3}(a) shows the first eleven elements of the JC model spectrum. Fig. \ref{fig:Fig3}(b) shows the Wigner function for the boson mode reduced density matrix for the ground state at the critical couplings on resonance.
Fig. \ref{fig:Fig3}(c) displays the first excited state at that critical crossing.

\section{Factorizable regime}
\label{sec:Sec5}

Let us construct an AR model that is {\em factorizable}, 
\begin{align}
    \frac{\hat{H}_{FAR}}{\hbar} = \frac{2}{2} \left\{  \hat{A} ,  \hat{A}^{\dagger} \right\},
\end{align}
in terms of the superposition of the superalgebra exchange operators, 
\begin{align}
    \hat{A} = \alpha_{0} + \alpha_{Q} \hat{Q}_{-} + \alpha_{R} \hat{R}_{-},
\end{align}
with complex constant parameters $\alpha_{j} \in \mathbb{C}$, $j=0, Q, R$. This leads to an explicit form, 
\begin{align}
    \begin{aligned}
        \frac{\hat{H}_{FAR}}{\hbar} =&~  \omega \hat{a}^{\dagger} \hat{a} + \frac{1}{2} \omega_{0} \hat{\sigma}_{z} + \lambda \left( e^{i \phi_{\lambda}} \hat{Q}_{+} + e^{-i \phi_{\lambda}} \hat{Q}_{-} \right) + \mu \left( e^{i \phi_{\mu}} \hat{R}_{+} + e^{-i \phi_{\mu}} \hat{R}_{-} \right) + \omega_{c},
    \end{aligned}
\end{align}
with effective parameters, 
\begin{align}
    \begin{array}{ll}
        \omega = \frac{1}{2} \left( \vert \alpha_{Q}\vert^{2} + \vert \alpha_{R}\vert^{2} \right), &~ 
        \omega_{0} = \frac{1}{2} \left( \vert \alpha_{Q}\vert^{2} - \vert \alpha_{R}\vert^{2} \right), \\
        \lambda = \vert \alpha_{0} \alpha_{Q} \vert, 
        &~ \phi_{\lambda} = \phi_{0} - \phi_{Q}, \\
        \mu = \vert \alpha_{0} \alpha_{R} \vert, 
        &~ \phi_{\mu} = \phi_{0} - \phi_{R}, \\ 
        \omega_{c} = \vert \alpha_{0} \vert^{2} + \frac{1}{2} \omega.
    \end{array}
\end{align}
Here we defined $\alpha_{j} = \vert \alpha_{j} \vert e^{i \phi_{j}}$, with $\alpha_{0}$ and $\alpha_{R}$ fulfilling the requirement, 
\begin{align}
    \begin{aligned}
        \Delta = \omega_{0} - \omega = \vert \alpha_{R} \vert^{2} = \frac{\mu^{2}}{\vert \alpha_{0} \vert^{2}}\cdot
    \end{aligned}
\end{align}
The latter implies that the detuning is proportional to the squared aJC coupling strength $\mu$ and inversely proportional to the  squared common factor between the JC and aJC coupling strengths $\vert\alpha_{0}\vert$.
Interestingly, a different way of writing this requirement, 
\begin{align}
    2 \omega \omega_{0} = \frac{1}{2}  \left( \vert \alpha_{Q}\vert^{4} - \vert \alpha_{R}\vert^{4} \right),
\end{align}
allows us to identify a reported exceptional regime \cite{Tomka2014} as the part of our general form for real coupling strengths.

\begin{figure}[ht]
\centering
\fbox{\includegraphics[scale=1]{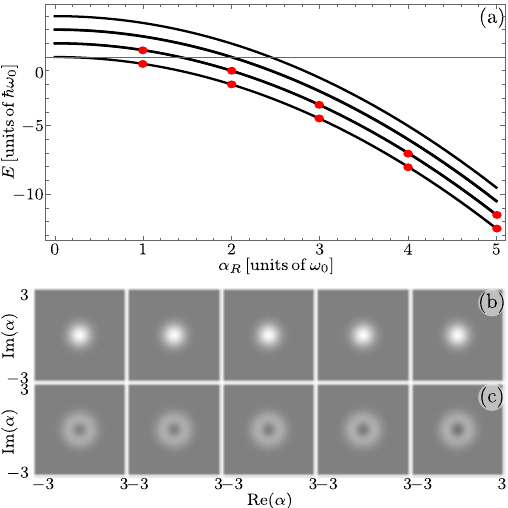}}
\caption{Same as Fig. \ref{fig:Fig1} but for the exceptional regime of the AR model with parameters $\omega = \omega_{0} - \vert \alpha_{R} \vert^{2}$, $\alpha_{0} = 0.01$, $\alpha_{Q} = \omega_{0}$. 
As there are no crossings in the spectrum, the points show arbitrary couplings $\alpha_{R} = \left\{1, 2, 3, 4, 5\right\} \omega_{0}$.}
\label{fig:Fig4}
\end{figure}

In our exceptional regime, where the AR model is factorizable in terms of the superalgebra exchange operators, the spectrum is, as shown in Fig. \ref{fig:Fig4}, equidistant, shows no crossings, and is double degenerate, with the exception of the unique lowest energy state. That is, it is isomorphic to the SUSY Hamiltonian, Eq.(\ref{eq:HQ}).

\section{Conclusion}
\label{sec:Sec6}

We reviewed the Jaynes--Cummings model from the Lie theory point of view, aiming to highlight the importance of an operator-based approach to quantum mechanics in order to create insight about physical systems using their abstract symmetries.
We favor an explicit approach to diagonalize the JC model to show the nuances required to construct its proper states and energies in the laboratory frame; in this case, the necessity to identify an parcel ground and excited subspaces in order to work with a representation of $su(2)$ for its diagonalization.
This help us to identify the well-known instabilities in the ground state of the JC model; that is, the critical couplings where the minimum energy becomes degenerate. 

For the sake of providing an explicit account, we reviewed the anti-Jaynes--Cummings model, which is isomorphic to the JC model up to a unitary rotation around the spin-half fermion mode.
Obviously, the spectral structure is identical to that of the JC model showing an unbroken SUSY that breaks at identical critical couplings. 
The alternate representation of the $u(1|1)$ superalgebra corresponding to this model will help deal with the anisotropic Rabi model.

Armed with the explicit representations for the two SUSY sectors corresponding to the JC and aJC model involved in the AR model, it was straightforward to identify a new sector provided by the $su(1,1)$ algebra related to single boson mode squeezing. 
We confirmed that the Rabi model is a singular limit of the AR model; that is, it is not possible to follow the squeezing approach as the squeezing parameter diverges when the JC and aJC coupling strengths become identical.
The AR model is mapped into a JC model plus parametric oscillator terms after transformation into the squeezed reference frame. 
Here, the logical step was to analyze the regime where the parametric drive is neglected and the effective model reduces to a JC model where our operator-based diagonalization approach is straightforward to find the spectral characterization of the model in this approximate regime.

Finally, we used the exchange operators of the SUSY sectors to build a factorizable AR (FAR) model in terms of the suporposition of these exchange operators. 
In this exceptional regime, the detuning between the frequencies of the fermion and boson mode needs to be proportional to the squared aJC coupling strength and inversely proportional to the common factor between the JC and aJC coupling strengths.
The energy spectrum of the FAR model shows an equally spacing and double degeneration but for an unique minimum energy ground state; that is, the characteristics of the SUSY Hamiltonian for an underlying $u(1\vert1)$ superalgebra.

We hope our manuscript provides a practical example of the value of the Lie theory operator-based approach in understanding the characteristics of physical systems and motivates its use beyond the mathematical physics community.

\section*{Funding}
 A.K. master studies funded by CONAHCYT \#1314608.  F.H.M.V. funded by CONAHCYT IxM \#551. 
\section*{Acknowledgments}	
B.M.R.L acknowledges fruitful discussion with Ricardo Guti\'errez J\'auregui.

	
\section*{References}

%

\end{document}